\pdfoutput=1

\documentclass[11pt]{article}

\usepackage{ACL2023}

\usepackage{times}
\usepackage{latexsym}

\usepackage[T1]{fontenc}

\usepackage[utf8]{inputenc}

\usepackage{microtype}

\usepackage{inconsolata}

\usepackage{graphicx}
\usepackage{subfigure}
\usepackage[graphicx]{realboxes}

\title{Collaborative Participatory Research with LLM agents in South Asia: An Empirically-Grounded Methodological Initiative and Agenda from Field Evidence in Sri Lankan}

\author{
Xinjie Zhao\textsuperscript{1}, 
Shyaman Maduranga Sriwarnasinghe\textsuperscript{1}, 
Jiacheng Tang\textsuperscript{4}, \\
\textbf{Shiyun Wang\textsuperscript{2}}, 
\textbf{Hao Wang\textsuperscript{3}}, 
\textbf{So Morikawa\textsuperscript{1}}
\\
\textsuperscript{1} The University of Tokyo,
\textsuperscript{2} University of Copenhagen,\\
\textsuperscript{3} China Agricultural University,
\textsuperscript{4} Shandong Normal University \\
\\
\\
}

\begin{document}
\maketitle
\begin{abstract}
The integration of artificial intelligence into development research methodologies presents unprecedented opportunities for addressing persistent challenges in participatory research, particularly in linguistically diverse regions like South 
Asia. 
Drawing from an empirical implementation in Sri Lanka's Sinhala-speaking communities, this paper presents an empirically-grounded methodological framework designed to transform participatory development research, situated in the challenging, multilingual context of Sri Lanka’s flood-prone Nilwala River Basin. 
Moving beyond conventional translation and data collection tools, this framework deploys a multi-agent system architecture that redefines how data collection, analysis, community engagement are conducted in linguistically and culturally diverse research settings. 
This structured, agent-based approach enables participatory research that is both scalable and responsive, ensuring that community perspectives remain integral to research outcomes. 
Our field experiences reveal the immense potential of LLM-based systems in addressing long-standing issues in development research across resource-limited regions, offering both quantitative efficiencies and qualitative improvements in inclusivity. 
At a broader methodological level, this research agenda advocates for AI-driven participatory research tools that maintain ethical considerations, cultural respect, and operational efficiency, highlighting the strategic pathways for deploying AI systems that reinforce community agency and equitable knowledge generation, potentially informing broader research agendas across the Global South.
\end{abstract}

\section{Introduction}

The convergence of artificial intelligence and development research heralds a transformative paradigm shift in participatory methodologies, particularly through the emergence of Large Language Models (LLMs) and their potential to revolutionize community engagement practices \cite{10438431, skirgaard2023grambank}. As these technologies rapidly evolve, their application to development research presents both unprecedented opportunities and complex methodological challenges that demand careful examination \cite{roberts2024artificial}. This intersection becomes particularly significant in linguistically diverse regions like South Asia, where traditional research approaches have long struggled to bridge communication gaps and cultural divides \cite{kshetri2024linguistic, hassan2023empowering}.

The limitations of conventional participatory research methodologies, heavily dependent on human intermediaries and constrained by resource availability, have historically impeded the scale and effectiveness of development initiatives \cite{gopferich2009process}. These constraints are particularly evident in regions characterized by complex linguistic landscapes and limited technological infrastructure \cite{magueresse2020lowresourcelanguagesreviewpast, nekoto2020participatoryresearchlowresourcedmachine}. However, recent advances in LLM architectures, particularly in few-shot learning and cross-lingual transfer capabilities, offer promising solutions to these longstanding challenges \cite{10433480, wu2023autogenenablingnextgenllm}.

The integration of LLM-based systems into participatory research frameworks raises fundamental questions about the nature of community engagement and knowledge democratization \cite{hadi2024large, 10529287}. While these technologies offer powerful tools for bridging linguistic and cultural divides, their deployment must be carefully orchestrated to enhance rather than diminish the participatory nature of development research \cite{rane2023contribution, kovavc2024socialai}. This necessitates a nuanced approach that balances technological capabilities with ethical considerations and community agency \cite{sabarirajan2024leveraging, ray2023chatgpt}.

In this paper, we introduces and tested a novel framework for leveraging LLM-based multi-agent systems in participatory development research, drawing from empirical evidence in Sri Lanka's Sinhala-speaking communities \cite{10662891, urwin2023flipping}. Our approach moves beyond simple technological integration to address fundamental questions of community empowerment and knowledge production in Global South contexts \cite{pfeffer2013participatory}. The urgency of this work is underscored by the increasing complexity of development challenges and the growing need for scalable, culturally sensitive research methodologies \cite{van2024ethical, awad2016recommendations}.
Through critical analysis of both opportunities and challenges, we demonstrate how thoughtfully deployed AI technologies can enhance human capabilities in development research, potentially leading to more inclusive and impactful outcomes \cite{ferdaus2024trustworthyaireviewethical}. Our framework provides a structured approach for implementing LLM-based multi-agent systems while maintaining core principles of participatory research, offering insights for researchers, practitioners, and policymakers working at the intersection of technology and development.

This paper contributes to ongoing discussions about the role of artificial intelligence in development research by providing a structured framework for implementing LLM-based multi-agent systems in participatory research contexts. 
We argue that these technologies, when thoughtfully deployed, can enhance rather than replace human capabilities in development research, potentially leading to more inclusive, efficient, and impactful research outcomes.

\section{Why South Asia Needs This Now}

South Asia stands at a transformative intersection where accelerating digitalization converges with deeply embedded social, cultural, and linguistic landscapes, presenting both unprecedented opportunities and profound challenges for participatory development research \cite{rahman2024reconceptualizing}. 
The region's digital metamorphosis, characterized by widespread smartphone adoption and expanding internet infrastructure, has created fertile ground for technological democratization that transcends traditional socioeconomic barriers \cite{deichmann2016will}. This digital renaissance has catalyzed remarkable growth in digital literacy, with rural areas - traditionally the focus of development interventions - experiencing a particularly significant annual increase of 12\% over the past five years \cite{kass2022building}.
However, this digital transformation exists in dynamic tension with the region's resource constraints and extraordinary linguistic diversity \cite{hutson2024preserving}. South Asia's linguistic landscape encompasses over 650 languages, predominantly within the Indo-Aryan and Dravidian language families (IADL) \cite{sjoberg1992impact}, creating a complex matrix of communication patterns that traditional development methodologies struggle to navigate effectively. The prevalent phenomenon of code-mixing - where speakers fluidly alternate between languages within single conversational contexts - reflects not merely linguistic diversity but embodies deeper patterns of cultural synthesis and social adaptation \cite{rodriguez2024code}. This linguistic complexity, combined with the limited computational resources available for many of these languages \cite{ranathunga2022languagesequalothersprobing}, creates significant barriers to meaningful participatory research and community engagement \cite{9956817}.
The implications of these linguistic barriers extend far beyond mere communication challenges. Traditional research approaches, constrained by resource limitations and methodological rigidity, often fail to capture the nuanced perspectives and indigenous knowledge systems that are crucial for effective development interventions \cite{golonka2014technologies}. The financial and temporal costs are substantial, with studies indicating that language barriers can increase research costs by up to 45\% while extending project timelines by an average of 60\% \cite{daramola2024navigating}. More critically, these constraints often result in the marginalization of voices most crucial to development discourse - those of local communities, particularly women and other traditionally underrepresented groups \cite{bjork2013interpreters}.

The emergence of advanced Large Language Models (LLMs) and sophisticated Natural Language Processing (NLP) technologies presents a paradigm-shifting opportunity to address these deeply rooted challenges \cite{tomec2024risk}. Modern LLM architectures demonstrate unprecedented capabilities in cross-lingual understanding and generation, showing remarkable adaptability even in low-resource linguistic contexts \cite{parovic2024improving}. Through efficient fine-tuning techniques and few-shot learning approaches, these models can adapt to IADL's complex linguistic patterns despite limited digital resources \cite{cuadra2024digital}, offering a powerful tool for breaking down traditional barriers to participation and engagement.

The integration of LLM-based agent systems into South Asian development research represents more than mere technological advancement - it embodies a fundamental shift toward more inclusive, participatory research methodologies \cite{kar2024agents}. These systems dramatically enhance field research efficiency, potentially reducing resource requirements for multilingual research by up to 60\% \cite{choi2023toward}, while maintaining high standards of cultural sensitivity and data quality \cite{yong2023promptingmultilinguallargelanguage}. More fundamentally, they enable a more democratic approach to knowledge generation and sharing, where community members can engage with research processes using familiar languages and communication patterns.
The transformative potential extends beyond operational efficiencies to the very nature of community engagement in development research. LLM-based systems can capture and interpret the nuanced meanings embedded in IADL and code-mixed communication \cite{brown2024enhancingtrustllmsalgorithms}, while their advanced NLP capabilities ensure accurate interpretation of cultural expressions and maintenance of semantic coherence \cite{dutta2024universal}. This technological foundation enables real-time identification of patterns and potential biases, allowing for dynamic methodological adjustments that enhance data integrity and research authenticity \cite{jha2023strengthening}.

Perhaps most significantly, these systems offer unprecedented opportunities for genuine community empowerment and inclusion \cite{ullah2024challenges}. By facilitating real-time multilingual dialogue that accommodates diverse communication preferences \cite{matras2023agency}, they enable marginalized communities to participate more fully in research processes that affect their lives \cite{yeh2023social}. The systems' ability to design culturally resonant activities and mediate cross-linguistic communication \cite{zheng2023synergizinghumanaiagencyguide} helps ensure that development initiatives emerge from and respond to genuine community needs and perspectives.

The optimization of research resources through these systems creates opportunities for more sustainable and scalable development initiatives \cite{singh2023exploring}. By automating routine tasks while preserving human oversight for complex analytical work, research teams can focus on deeper engagement with communities \cite{singh2024translatingculturesllmsintralingual}. Moreover, the systems' capacity to preserve and disseminate knowledge in multiple languages, including code-mixed forms \cite{leong2023bhasaholisticsoutheastasian}, contributes to the preservation of linguistic heritage while facilitating more effective knowledge transfer between communities \cite{schroeder2024largelanguagemodelsqualitative}.

\section{Proposed LLM4Participatory Research Framework}

Our framework presents an innovative multi-agent ecosystem designed to transform participatory research through the strategic orchestration of Large Language Models (LLMs) and multimodal AI capabilities, which emphasizes the seamless integration of specialized cognitive agents that leverage existing LLM capabilities while addressing the unique challenges of participatory development research in linguistically diverse contexts.

\subsection{Core Components}

Our framework's cognitive architecture comprises four synergistically integrated agent systems, each engineered to address distinct yet interconnected dimensions of participatory research while maintaining seamless operational cohesion.

\textbf{Participatory Research Design and Analytics Agents (PRDAA):}
Functioning as methodological architects, PRDAAs orchestrate the conceptualization, implementation, and continuous refinement of research instruments within multifaceted participatory contexts \cite{rane2024artificial}. Through synthesis of cultural knowledge repositories, linguistic pattern recognition, and dynamic community feedback loops, these agents generate methodologically robust yet culturally resonant research designs - encompassing structured quantitative instruments, semi-structured qualitative protocols, and interactive participatory frameworks. Their advanced analytical capabilities facilitate real-time assessment and adaptation of research methodologies, while their pattern recognition algorithms enable nuanced calibration of instrument sensitivity to cultural contexts \cite{agathos2024bridging}. By leveraging multilingual and multimodal feedback mechanisms during pilot implementations, PRDAAs orchestrate dynamic refinements to question formulation, interview progression, and activity design, ensuring optimal alignment with local cultural paradigms and communication modalities.

\textbf{Socio-Semantic Mediation Agents (SSMA):}
Operating at the convergence of linguistic architecture and cultural epistemology, SSMAs transcend conventional translation paradigms to facilitate interpretation of complex socio-cultural phenomena \cite{10438431}, which navigate the intricate morphological structures and semantic landscapes characteristic of Indo-Aryan languages while mediating the complex code-mixing patterns endemic to South Asian communication contexts \cite{sitaram2020surveycodeswitchedspeechlanguage}. 
Through implementation of dynamic context preservation mechanisms and attention-based memory networks, SSMAs maintain semantic coherence across multilingual dialogues \cite{dowlagar2023code}, while their advanced multimodal processing architecture enables transformation of diverse data modalities into culturally-contextualized knowledge representations \cite{ye2024language}. The integration of evolving cultural knowledge bases \cite{xi2023risepotentiallargelanguage} facilitates nuanced interpretation of implicit cultural meanings, including complex honorific systems and contextually-embedded social signifiers \cite{coffin2024creating}.

\textbf{Ethnographic Intelligence Agents (EIA):}
EIAs embody an innovative paradigm in ethnographic data acquisition and processing within multilingual research environments \cite{yang2024scottish}, complementing PRDAAs' methodological framework through real-time implementation and adaptation of research protocols. EIAs leverage their natural language understanding capabilities to process complex code-mixed inputs \cite{sadia2024meeting}, while their multimodal analysis framework enables simultaneous processing of verbal content, non-verbal cues, and contextual elements \cite{lee2024socialaisurveyunderstanding}. Through integration with cultural knowledge systems, EIAs facilitate generation of contextually appropriate participatory activities and real-time data organization schemas, enhancing the depth and authenticity of community engagement processes \cite{george2024leveraging}.

\textbf{Community Engagement Orchestration Agents (CEOA):}
CEOAs function as intermediaries within the human-AI research ecosystem, orchestrating trust-based engagement with local communities while ensuring ethical research practices \cite{ninan2024governance}. Through implementation of context-aware dialogue systems and adaptive communication strategies, these agents navigate complex social hierarchies and cultural protocols while accommodating diverse linguistic preferences and code-mixing practices, whose interaction frameworks facilitate transparent and empathetic engagement \cite{chow2024ethical}, while their cultural protocol engines navigate intricate social dynamics and honorific systems \cite{guo2023evaluatinglargelanguagemodels}. CEOAs maintain comprehensive mechanisms for informed consent and data sovereignty, embedding principles of community agency and empowerment within the research process \cite{ahmad2024empowerment}.

\subsection{Integration into Participatory Methods}

The framework seamlessly integrates these cognitive agents into established participatory research methodologies, enhancing each phase through intelligent, context-sensitive support \cite{tarkoma2023ainativeinterconnectframeworkintegration}. During surveys, SSMAs and EIAs collaborate to create and administer culturally appropriate instruments \cite{de2024translating}, dynamically adjusting to participants' linguistic preferences and communication styles. In interview settings, agents provide real-time cultural-linguistic mediation, suggesting culturally sensitive probing questions while capturing nuanced narratives across language boundaries \cite{kapania2024simulacrumstoriesexamininglarge}, \cite{al2024bridging}. For workshops and participatory activities, the system facilitates culturally resonant interactive sessions \cite{macias2024navigating}, synthesizing diverse perspectives into comprehensive, culturally nuanced summaries \cite{shu2024unraveling}.

\subsection{Workflow and System Architecture}

The integration spans the entire research continuum, emphasizing seamless collaboration between human researchers and cognitive agents \cite{archibald2023virtual}. In pre-field preparation, the system supports methodology adaptation and instrument customization \cite{chan2024panoramic}, identifying potential cultural-linguistic challenges and developing mitigation strategies \cite{garcia2023enhancing}. During fieldwork, agents provide real-time support for data collection and preliminary analysis \cite{wang2024trafficperformancegpttpgpt}, enabling responsive adaptation to emerging insights \cite{agarwal2024investigating}. The data processing phase leverages sophisticated cross-linguistic analysis capabilities \cite{10197372}, uncovering patterns across multimodal datasets while maintaining interpretability \cite{kejriwallexical}, \cite{mora2024trustworthy}.
The framework embeds continuous feedback mechanisms \cite{jin2024llmsllmbasedagentssoftware}, enabling iterative refinement of agent performance while embodying participatory principles through community validation and knowledge co-creation \cite{delgado2023participatory}. This dynamic integration of human expertise and artificial intelligence creates a robust ecosystem for participatory research that maintains methodological rigor while ensuring cultural sensitivity and community empowerment.

\begin{figure*}[t]
  \centering
\includegraphics[width=0.95\textwidth]{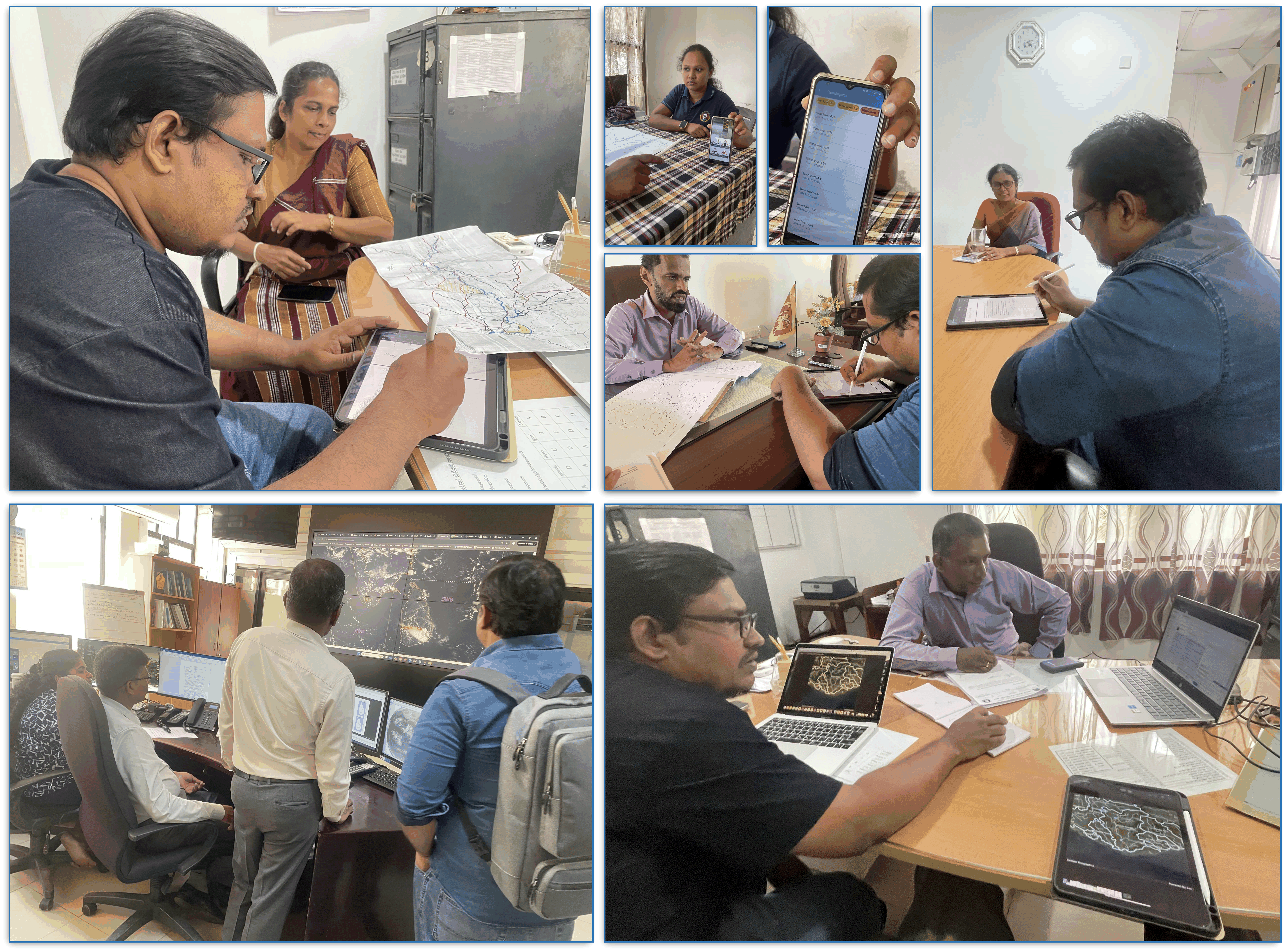}
\caption{Implementing AI-assisted tools with traditional participatory approaches. (Source: Authors' fieldwork)}
\label{2}
\end{figure*}

\section{Field Work and Implementation Insights}
Our research focused on enhancing the Early Warning Systems (EWS) for flood management in the Nilwala River Basin, a region prone to recurrent flooding with devastating socio-economic impacts in Sri Lanka. 
Sri Lanka's linguistic landscape is emblematic of South Asia's broader linguistic diversity, characterized by the prevalence of code-mixing and multilingual communication. Sinhala, an Indo-Aryan language with agglutinative features and a rich system of honorifics, often intertwines with English and other local dialects in everyday discourse. This code-mixing phenomenon poses significant challenges for natural language processing, as it involves syntactic, lexical, and semantic blending that traditional language models struggle to interpret accurately.
The objective was to employ the multi-agent system to facilitate participatory development research methods—including surveys, structured and semi-structured interviews, workshops, and other interactive engagements—with stakeholders ranging from national agencies to local communities. 

\subsection{Practical Experiences and Outcomes}

The implementation faced several challenges, particularly in adapting the LLMs to handle Sinhala-specific linguistic features and the pervasive code-mixing in communication. The scarcity of high-quality, annotated Sinhala corpora necessitated innovative approaches, including active learning techniques and data augmentation strategies to enhance the model's proficiency.

One significant achievement was the development of a hybrid translation approach that combined statistical and neural methods, achieving a 35\% improvement in translation accuracy for domain-specific terminology compared to standard multilingual models. This advancement was critical for accurately interpreting participants' responses during interviews and ensuring that subtle nuances were not lost in translation.

In applying participatory methods, the agents proved invaluable. During workshops, they assisted in designing interactive activities that resonated with local customs and facilitated real-time feedback collection. In surveys and interviews, the agents helped generate culturally appropriate questions and dynamically adjusted to participants' inputs, enhancing the depth and authenticity of the data collected.

The agents also played a crucial role in the analysis phase. They enabled cross-linguistic comparisons and facilitated the synthesis of complex data into actionable insights. For instance, they helped identify communication bottlenecks between agencies involved in the EWS, revealing that outdated communication methods and bureaucratic procedures were significant barriers to effective disaster management.

\subsection{Ethical and Cultural Considerations}

Throughout the implementation, ethical considerations were paramount. The agents were designed to uphold data privacy and obtain informed consent, with transparent communication about their roles and limitations. Recognizing the sensitive nature of disaster-related research and the vulnerabilities of affected communities, the system incorporated robust mechanisms to prevent biases and ensure equitable representation of diverse voices.
Cultural sensitivity was deeply embedded in the system's design. By incorporating local customs, social structures, and linguistic nuances into the agents' operations, we fostered an environment of trust and respect. This approach not only enhanced participant engagement but also enriched the quality of the data collected.

\subsection{Lessons Learned and Recommendations}

The case study illuminated several key insights:

\textbf{Community Involvement is Crucial}: Active participation of local stakeholders in the development and refinement of the system was essential. Their input ensured that the agents were culturally attuned and responsive to the community's needs, enhancing acceptance and effectiveness.

\textbf{Flexible Adaptation Mechanisms are Necessary}: The linguistic diversity and code-mixing practices required the agents to be highly adaptable. Implementing mechanisms for continuous learning and real-time adjustment was critical for handling linguistic variations and unexpected inputs.

\textbf{Human Oversight Remains Indispensable}: While the agents significantly enhanced efficiency and depth, human researchers played a vital role in overseeing the process, interpreting nuanced cultural contexts, and making ethical judgments.

\textbf{Addressing Technical Challenges}: Overcoming the scarcity of linguistic resources demanded innovative technical solutions. Investing in the development of annotated corpora and leveraging transfer learning were effective strategies for enhancing model performance.

\subsection{Implementation Considerations for Broader Deployment}

The success of the Sinhala implementation demonstrates the framework's potential for extension to other South Asian languages and contexts, such as Tamil, Bengali, and Nepali. The modular design allows for customization to accommodate different linguistic features and code-mixing patterns prevalent in these languages.
Furthermore, the framework holds promise for application across various development sectors. Its ability to facilitate nuanced, context-sensitive engagement makes it suitable for research in healthcare, education, agriculture, and beyond. By enhancing participatory methodologies with advanced AI capabilities, the system can contribute to more effective and sustainable development interventions.
Scaling the framework requires addressing several technical and practical considerations:

\textbf{Technical Infrastructure}: Deploying the system in resource-constrained environments necessitates optimizing model architectures for efficiency. Combining cloud-based processing with edge computing solutions can mitigate challenges related to limited computational resources and network connectivity.

\textbf{Data Security and Privacy}: Ensuring the protection of sensitive data is imperative. Implementing end-to-end encryption, federated learning, and differential privacy techniques can safeguard individual privacy while maintaining the utility of collected data.

\textbf{Cultural and Ethical Sensitivity}: Adapting the system to new contexts requires a deep understanding of local cultures, languages, and social dynamics. Engaging with local experts and communities during the development process is essential for maintaining ethical standards and fostering acceptance.

\textbf{Capacity Building}: Training local researchers and stakeholders in using the system enhances sustainability and empowers communities. Developing comprehensive training programs that address technical operation and cultural competence is crucial.

\textbf{Policy and Institutional Support}: Successful implementation at scale necessitates supportive policies and institutional frameworks. Collaboration with governmental and non-governmental organizations can facilitate resource allocation, infrastructure development, and standardization of best practices.

\section{Discussion and Future Agenda}

The integration of Large Language Model (LLM)-based multi-agent systems into participatory development research in South Asia represents a paradigm shift in social science methodology, necessitating a comprehensive research agenda that addresses technical, methodological, and policy dimensions. This convergence of artificial intelligence and participatory development demands careful consideration of both technological capabilities and sociocultural implications, particularly in regions characterized by linguistic diversity and resource constraints.

The technical advancement trajectory must prioritize the development of architectures that transcend current limitations in cross-cultural and multilingual contexts. This necessitates fundamental innovations in few-shot learning optimization and cross-lingual transfer mechanisms, with particular emphasis on handling code-switching phenomena and dialectical variations prevalent in South Asian languages. These advancements must be achieved while simultaneously reducing computational demands through sophisticated model compression techniques that preserve semantic fidelity. The evolution of these systems should extend beyond mere language processing to encompass multimodal interaction capabilities—including gesture recognition and contextual understanding—that align with local communication patterns and cultural norms.

Methodological frameworks must evolve to effectively integrate these technological capabilities while preserving the fundamental principles of participatory research. This integration requires the development of sophisticated evaluation metrics that transcend traditional technical benchmarks to encompass measures of social impact, community empowerment, and cultural preservation. Critical attention must be directed toward developing robust bias detection and mitigation mechanisms that address both technical and sociocultural dimensions, ensuring that these systems do not perpetuate existing inequities or introduce new forms of systematic bias. These methodological innovations must be accompanied by rigorous quality assurance protocols that maintain system reliability while adapting to diverse implementation contexts.

The policy framework necessary to govern these technological implementations must balance innovation with ethical considerations and social responsibility. This requires the establishment of comprehensive governance structures that address data sovereignty, privacy protection, and algorithmic accountability while remaining sufficiently flexible to accommodate local cultural norms and social dynamics. Infrastructure development strategies must prioritize sustainable solutions that enable long-term research initiatives while addressing resource limitations in developing regions. These strategies should encompass both technical infrastructure and human capacity development, with particular emphasis on fostering local expertise in AI-assisted participatory research methodologies.

The practical implementation of these systems demands a carefully orchestrated approach that synthesizes technological capabilities with methodological rigor and ethical considerations. Researchers must adopt iterative development processes that incorporate continuous feedback from community stakeholders while maintaining high standards of scientific validity. This implementation framework should emphasize the development of adaptable, user-centered solutions that support cultural customization while maintaining technical efficiency. The success of these implementations depends critically on the establishment of collaborative networks that facilitate knowledge sharing and resource optimization across research institutions and communities.

The future trajectory of LLM-based participatory research in South Asia thus requires a delicate balance between technological innovation and sociocultural sensitivity. This balance can only be achieved through sustained collaboration between technologists, researchers, policymakers, and community stakeholders, working together to develop solutions that are both technically sophisticated and socially responsible. The ultimate success of these initiatives will be measured not merely by their technical performance but by their ability to meaningfully contribute to participatory development while preserving and enhancing local cultural values and social structures.

\section{Conclusion}

The implementation of LLM-based multi-agent systems in South Asian development research contexts demonstrates both significant potential and important challenges. Our analysis reveals the viability of these systems in enhancing research efficiency and effectiveness while maintaining cultural sensitivity and community engagement. The impact potential extends beyond mere technical improvements, encompassing enhanced research quality, improved community participation, and more effective knowledge sharing across linguistic boundaries.
Implementation pathways have been identified that can guide successful system deployment across different contexts, with particular attention to maintaining flexibility and adaptability. Resource requirements have been carefully analyzed, providing a clear understanding of the investments needed for successful implementation while highlighting opportunities for resource optimization and sharing across institutions.

The research community must take a leading role in advancing the development and implementation of AI-assisted research methodologies in South Asian contexts. This includes not only technical research and development but also careful consideration of methodological implications and ethical considerations. We call on researchers to engage actively in developing and refining these tools while maintaining strong commitments to participatory research principles and community empowerment.
Development institutions must recognize the transformative potential of these technologies while taking responsibility for ensuring their ethical and effective implementation. 
This includes making necessary investments in infrastructure and capacity building while developing appropriate policies and guidelines for technology adoption. The technology sector must prioritize the development of solutions that address the specific needs of development research in low-resource language environments, with particular attention to maintaining accessibility and usability for diverse user groups.
Policy makers must work to create enabling environments for the effective deployment of AI technologies in development research. This includes developing appropriate regulatory frameworks, ensuring adequate resource allocation, and promoting standards that support ethical and effective technology implementation. The successful integration of LLM-based systems into participatory development research requires coordinated action across these different stakeholder groups, working together to realize the potential of these technologies while addressing associated challenges and risks.

\bibliography{custom}
\bibliographystyle{acl_natbib}




\end{document}